# Design and Field Implementation of Blockchain Based Renewable Energy Trading in Residential Communities

Shivam Saxena & Hany Farag
*Dept. of Electrical Engineering and Computer Science*
York University
Toronto, Canada
{shivam,hefarag}@cse.yorku.ca

Aidan Brookson
*Sustainable Tech. Eval. Program*
Torono and Region
Conservation Authority
Toronto, Canada
abrookson@trca.on.ca

Hjalmar Turesson & Henry Kim
*Schulich School of Business and Management*
York University
Toronto, Canada
{hturesson,hkim}@schulich.yorku.ca

*Abstract*—This paper proposes a peer to peer (P2P), blockchain based energy trading market platform for residential communities with the objective of reducing overall community peak demand and household electricity bills. Smart homes within the community place energy bids for its available distributed energy resources (DERs) for each discrete trading period during a day, and a double auction mechanism is used to clear the market and compute the market clearing price (MCP). The marketplace is implemented on a permissioned blockchain infrastructure, where bids are stored to the immutable ledger and smart contracts are used to implement the MCP calculation and award service contracts to all winning bids. Utilizing the blockchain obviates the need for a trusted, centralized auctioneer, and eliminates vulnerability to a single point of failure. Simulation results show that the platform enables a community peak demand reduction of 46%, as well as a weekly savings of 6%. The platform is also tested at a real-world Canadian microgrid using the Hyperledger Fabric blockchain framework, to show the end to end connectivity of smart home DERs to the platform.

## I. Introduction

Motivated by environmental concerns related to climate change and associated financial incentives, homeowners in residential communities are shifting towards procuring locally deployed distributed energy resources (DERs) that seek to maximally utilize clean, renewable energy to accomplish their respective tasks [1]. Within residential communities, these DERs typically include: rooftop photovoltaic arrays (PV), plug-in electric vehicles (EV), smart thermostats (ST), as well as battery energy storage systems (BESS) [2]. These DERs offer many tangible benefits to the community, including increased energy efficiency, reduction of peak demand, increased resiliency from outages in the main grid, as well as a decreased carbon footprint.

However, these DERs can have unintended negative consequences if left uncontrolled. Previous work has investigated the negative impact of uncontrolled EV charging leading to overloading of local transformers, as well as uncontrolled PV generation leading to overvoltage violations [3]. The addition of DERs to a home has indeed resulted in the vision of *smart* homes, however, there is a fundamental need for additional mechanisms that will coordinate and align the operation of smart home DERs to mitigate the aforementioned issues. Furthermore, since the DERs are owned by individual home owners within the community and not the community or utility, the mechanism must lend itself towards incentivized participation rather than unilateral control.

An emerging mechanism that aligns well within this concept is peer to peer (P2P) energy trading, where homeowners (analogous to peers in this context) can utilize their DERs to trade renewable energy amongst neighbors. The role of a traditional homeowner as an end consumer is, therefore, transformed into a *prosumer* that is capable of buying and selling electricity at its discretion. The trading process is facilitated by a virtual energy marketplace by enabling home owners to place energy "bids" for each DER per each discrete market interval during the day, and a market clearing process is used to determine whether the DER wins its submitted bid and operates at its preferred setting [4]. As such, P2P energy trading systems inherently balance local power mismatches, while also opening up new revenue streams for homeowners [5].

Yet, there are three major concerns with conventional P2P energy trading platforms that revolve around auditability, privacy, and security [6]. In a decentralized P2P system, it is challenging to verify the correctness of energy transactions amongst peers without access to the peers historical records. However, if historical records of peers are indeed made available, this raises severe privacy concerns that would significantly inhibit peer participation [7]. An alternative solution would be to use a trusted, central intermediary to verify all peer transactions, however, this exposes a single point of failure and renders the system insecure. Given these factors, there exists a major trust issue with disparate peers exchanging in energy trades over an unsecured, untrusted platform.

Recently, blockchain technology has been used to address the aforementioned challenges in P2P environments. A blockchain is a type of distributed ledger, where each peer maintains a local copy of the ledger and participates in a consensus process to verify all transactions made by all peers. Transactions submitted to the blockchain during a given time period are first encrypted to hide the identity of peers associated with the transaction, collected in a discrete block of data, verified by peers against a set of rules that the network is governed by, and then appended to the end of the ledger in a tamper-proof fashion. Transactions are generated and verified by smart contracts, which are software applications that are deployed to the blockchain and auto-execute based on the state of the ledger. Given these set of properties, blockchains are used as the transactive layer in P2P energy trading because they i) obviate the need for a central intermediary to facilitate/verify all transactions, ii) protect the privacy of peer transactions using specialized encryption techniques, and iii) do not expose a single point of failure. An example of a blockchain based P2P energy trading system can be seen in Fig. 1, where each smart home is a peer on the blockchain network, and utilizes the ledger to store energy measurements and details of trading transactions. Smart contracts auto-execute the business logic of clearing the market and administering the market in a trustless manner.

Blockchain based energy trading is receiving a great deal of attention in practice and literature [8]–[10]. The aforementioned work, however, uses a public blockchain implementation (mainly Ethereum and Bitcoin) that is open for the public to join, and requires every single participant to verify all transactions in return for incentives (referred to as mining). As such, public blockchains have significant scalability issues, as well as concerns with the sheer amount of energy the mining process consumes. For example, the energy consumed from Bitcoin mining in the year 2017 was equivalent to the annual energy consumption of Ireland, which is in excess of 30 TWh [11]. On the other hand, permissioned blockchains are invitation-only, and only require a subset of participants to verify transactions, leading to higher transaction throughput [12].

Consequently, this paper proposes a permissioned blockchain implementation for a P2P energy trading system for residential communities, where individual home owners are able to place energy bids for their DERs on the blockchain ledger for discrete time intervals. After all bids are collected, the market is cleared via a double auction method that is implemented as a smart contract, and individual DERs are sent control signals that determine their control setting for the specified market interval. Two sets of experimental results are carried out to test the efficacy of the proposed system. The first set of results are based on a single community of eight homes using real-world data, while the second set of results are of a real-world implementation within a Canadian microgrid of 4 homes. The simulated results show that the proposed system is able to reduce peak demand by 46% and reduce electricity bills for the community by 6% per week.

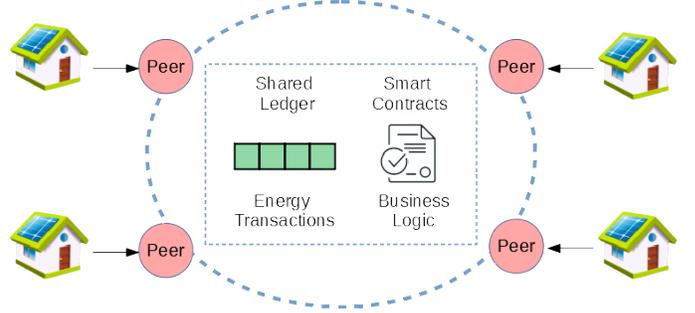

Fig. 1. An example blockchain P2P energy trading network.

The organization of the paper is as follows. Section II presents the mathematical formulation of the DERs, while Section III covers various bidding strategies used by the home owners and details the market clearing price algorithm. Section IV covers the blockchain implementation of the system, Section V provides detail on the experimental results, while Section VI concludes the paper and summarizes it's main contribution.

## II. MODELING OF SMART HOME DERs

Each smart home, designated by subscript $m$, may possess any combination of DERs, designated by subscript $n$. The following subsections describe the mathematical modeling of PV, BESS and EV units. The modelling of the ST can be found in [13], and is left out due to the length of the heat transfer equations.

### A. BESSs and EVs

The mathematical modeling for BESSs and EVs is quite similar and has been combined into one section to save space. The main constraints for BESS/EVs are that the power requirement should remain within the maximum and minimum limits of its onboard inverter, and that its state of charge (SoC) should remain within the recommended manufacturer's limits. One additional constraint for an EV is that it should reach a specified SoC at a certain time before a specified departure time. These three constraints are shown in (1)-(3), while the equation for the calculation of the current SoC is shown in (4). It should be noted that the terms BESS and EVs can be used interchangeably in this subsection, except for in (3), and that for the purposes of this manuscript, the EV is assumed to have only charging capability.

$$P_{BESS,m,n}^{MIN} \leq P_{BESS,m,n}(t) \leq P_{BESS,m,n}^{MAX} \quad (1)$$

$$SoC_{BESS,m,n}^{MIN} \leq SoC_{BESS,m,n}(t) \leq SoC_{BESS,m,n}^{MAX} \quad (2)$$

$$SoC_{EV,m,n}^{Dep}(t+\omega) \geq SoC_{EV,m,n}^{Req}(t) \quad (3)$$

$$SoC_{BESS,m,n}(t+1) = SoC_{BESS,m,n}(t)+$$
$$\left(\chi_{m,n} \cdot \eta_{m,n} + \frac{(1-\chi_{m,n})}{\eta_{m,n}}\right) P_{BESS,m,n}(t) \quad (4)$$

where, $\{P_{BESS}^{MIN}, P_{BESS}^{MAX}\}$ are the minimum and maximum power limits, $P_{BESS}(t)$ is the instantaneous power requirement, $\{SoC_{BESS}^{MIN}, SoC_{BESS}^{MAX}\}$ are the minimum and maximum SoC limits, $SoC_{BESS}(t)$ is the current SoC, $\{SoC_{EV}^{Dep}, SoC^{Req}(t)\}$ are the SoC at departure time and the SoC required before departure time, respectively, $\chi$ is a binary variable that represents 1 for charging mode and 0 otherwise, and $\eta$ is the charging/discharging efficiency.

### B. PV

The PV system consists of a PV array that generates direct current (DC), and is coupled with an inverter that converts the DC to alternating current (AC). The DC power generated from the PV array can be found as follows [14],

$$P_{PV,m,n}^{DC} = P_{PV,m,n}^{RT} IRR(t) F_T(T_{m.n})(t) \quad (5)$$

where, $P_{PV}^{DC}$ is the DC power generated by the PV array, $P_{PV}^{RT}$ is the nameplate rating of the PV array, $IRR$ is the current level of irradiance in kW/m$^2$, T is the current temperature in °C, and $F_T(T)$ is an interpolated temperature factor that can be found in [14]. The AC power output of the PV system can be found by multiplying the efficiency of the inverter with the DC power output as follows

$$P_{PV,m,n}(t) = P_{PV,m,n}^{DC}(t)\psi_{m,n} \quad (6)$$

where, $P_{PV,m,n}$ is the final AC output power of the PV system, and $\psi_{m,n}$ is the inverter efficiency that can be interpolated using methods and data found in [14].

### III. MARKETPLACE DESIGN

#### A. Review of Conventional MCP Calculation

The conventional strategy of computing the MCP of an electricity market is via a double auction, where potential sellers (DERs) and buyers (loads) of energy simultaneously submit energy bids to an auctioneer for discrete time slots during the day [15]. The auctioneer arranges the generator bids in ascending order of price, does the reverse for the load bids, and computes the intersection between the two resultant curves to find the MCP. For generators, bids that are below and to the left of the MCP are granted, and for loads, bids that are above and to the left of the MCP are granted. Formulating the MCP in this way results in the merit-order effect, where the highest bids of energy demand are satisfied by the most inexpensive generators. This is particularly true for markets with high penetrations of PV energy, since PV systems have very little production cost, and can therefore be used to accommodate more demand at lower prices [16]. An example bidding process is illustrated in Fig 2, where the MCP is shown for a system with limited (MCP1) and high levels of DER penetration (MCP2), respectively. It can be seen that in the second scenario, the merit order effect is greatly enhanced due to the greater influence of DER generation, resulting in a lower MCP ($0.07 versus $0.11) and greater system demand being able to be serviced (1.8 kWh versus 0.8 kWh).

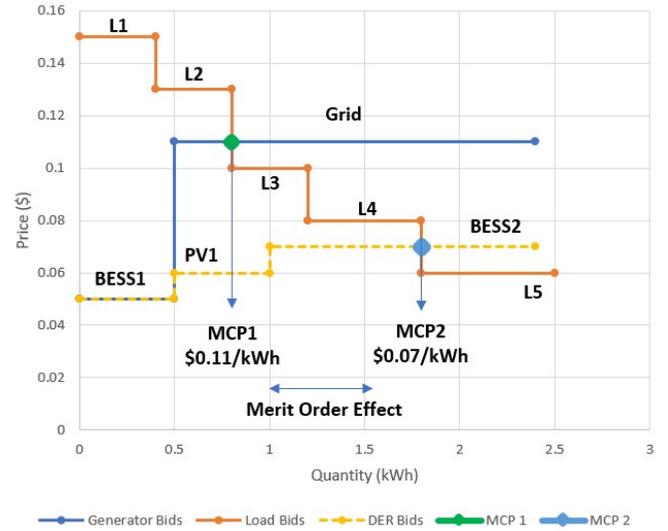

Fig. 2. Merit order effect enhanced when more DERs submit energy bids.

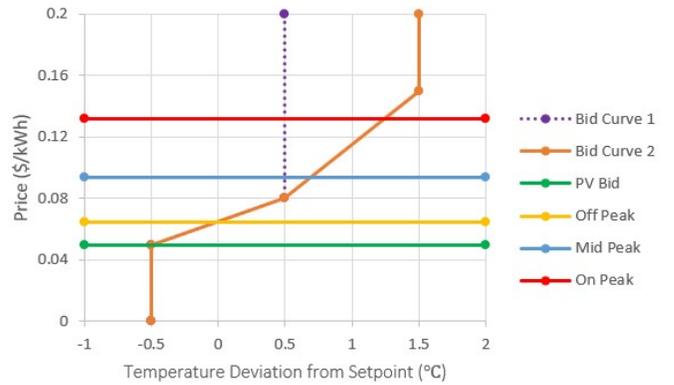

Fig. 3. Example bid curves of Smart Thermostats.

#### B. Proposed Bidding Strategies for Controllable DERs

In this paper, bidding strategies are defined for the controllable DERs, which are STs, BESSs, and EVs. PVs are assumed to have static bids as shown in Fig. 2, and are not discussed within this paper. In general, an energy bid is a reflection of how much a DER owner is willing to pay for the sacrifice of something that holds value, which can be modeled by a bid curve. In the case of STs, the item of value is the flexibility of thermal comfort, whereas for EVs, the item of value is the flexibility of the desired SoC dictated by its arrival and departure times. BESSs, by virtue of being able to both consume and generate energy, are able to formulate their strategy based on the level of SoC and opportunities to generate revenue depending on the current price of electricity. Generalizing the above, two bidding strategies are defined as *selfish*, and *helpful*, where a selfish bidding strategy tends to produce inflexible bid curves, and a helpful bidding strategy tends to produce less steep, flexible bid curves that are willing to sacrifice value in return for financial gain. It is worthwhile mentioning that the more selfish a bid curve is for a particular DER, the higher its energy demand and cost is for the homeowner, and the less impact the DER can have in participating in initiatives to reduce peak demand.

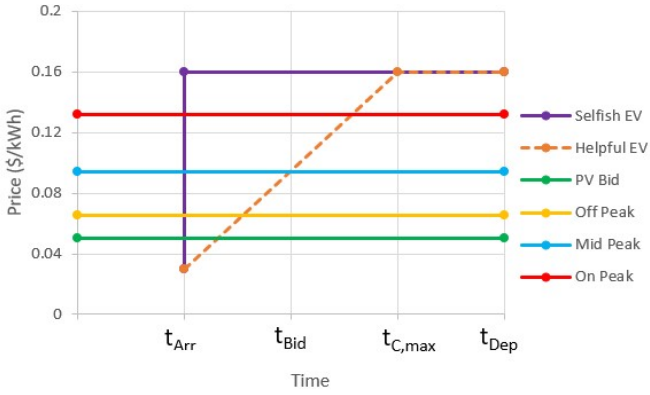

Fig. 4. Example bid curves of selfish and helpful EVs.

For example, two ST bid curves are depicted in Fig 3, where the bid curve symbolizes the incremental price a ST is willing to pay as a function of thermal discomfort, that is, as the deviation from the desired setpoint increases. Typical distribution system energy prices (time of use - TOU), along with a sample PV bid are also depicted in the figure, where the intersection of the ST bid curve with each subsequent supply curve represents the price that the ST is willing to pay per unit of temperature deviation from the setpoint. In this example, both bid curves become more steep as the deviation from the desired setpoint increases. However, it can be seen that bid curve 1 is much more steep than bid curve 2, and is willing to pay any price for cooling energy when the temperature deviation reaches 0.5 °C. As such, bid curve 1 is classified as being more selfish than bid curve 2.

An example of selfish/helpful bidding strategy for EVs is shown in Fig. 4, where relevant parameters of time are defined as $t_{Arr}$ (the home arrival time of the EV), $t_{Bid}$ (the time at which the EV places a bid), $t_{C,max}$ (the maximum time the EV can wait before it must charge at full power to reach a desired SoC based on the a time of departure), and $t_{Dep}$ (the departure time of the EV). The selfish EV bid curve shows that it will have the EV charge instantaneously upon $t_{Arr}$ and is willing to pay any electricity rate to charge. On the other hand, the helpful EV bid curve is more flexible, willing to bid in the market by ramping its energy demand until $t_{C,max}$, where its bid curve converges with the selfish bid curve since it will need maximum charging power to the desired SoC before $t_{Dep}$. The $t_{C,max}$ per EV can be calculated as,

$$t_{C,max} = \left[\frac{SoC_{EV}^{MAX} - SoC_{EV}(t)}{\eta \cdot P_{EV}^{MAX}}\right] \quad (7)$$

where the subscript $\{m,n\}$ is dropped for brevity.

The selfish/helpful bids for BESS revolve around financial incentive. As such, a selfish BESS will charge only in off-peak periods, or when there is excess PV energy available, and attempt to discharge at on-peak periods to gain maximum revenue. On the other hand, a helpful BESS will look to charge only when there is excess PV energy available, and discharge when its local load exceeds its local demand .

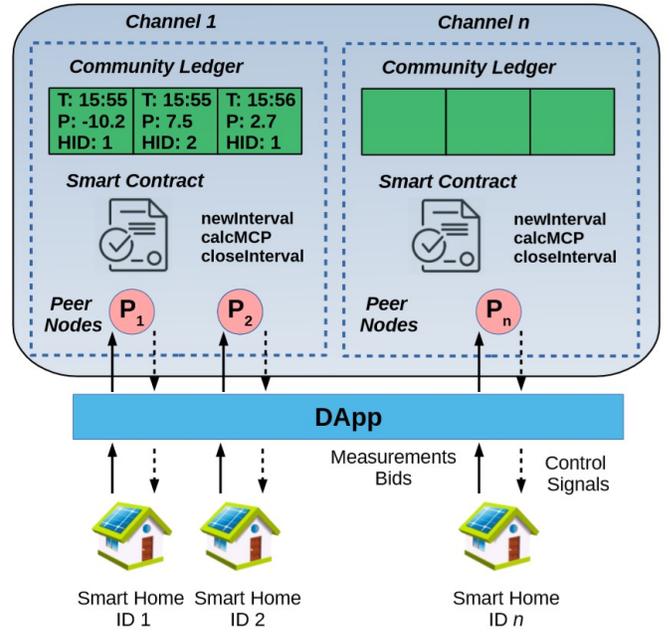

Fig. 5. Architecture of proposed blockchain based energy trading system.

## IV. PERMISSIONED BLOCKCHAIN IMPLEMENTATION

The proposed system is implemented using Hyperledger Fabric (HLF), which is an enterprise-grade permissioned blockchain framework. Unlike public blockchain implementations, HLF allows the overall system to be segmented into private channels, where peers may conduct transactions that are hidden from other channels. Each channel is designated its own ledger, in which the peers are responsible for maintaining the ledger state. Segmentation allows for better scalability and abstraction, since all peers are not required to validate all transactions throughout the entire system. A high-level architectural block diagram of the proposed system is illustrated in Fig. 5, where communities are segregated into their own channel, each self-governed by the peer nodes of the smart homes in the channel. A description of the system components are given below, in context of the proposed system:

• **Peer Node**: A node on the blockchain network that retains a copy of the ledger and participates in the consensus/verification process. A node within the energy trading platform would be an instance of a smart home.

• **DApp**: The front-end application that allows a smart home to interact with the blockchain, including placing bids, monitoring bid status, submitted energy measurements, and facilitating automated control of DERs.

• **Ledger**: Decentralized database that stores shared system data, including all energy measurements and energy bids for

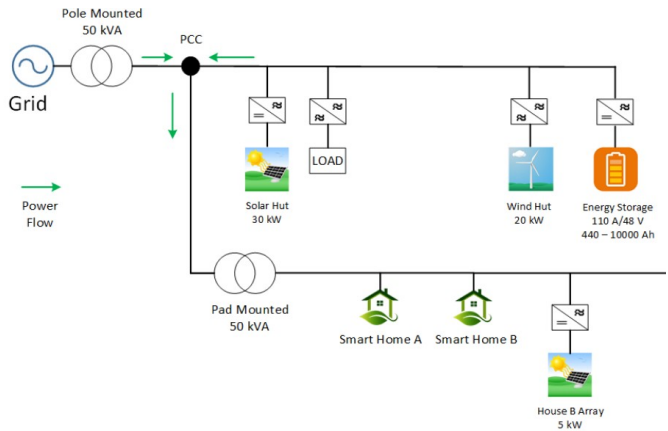

Fig. 6. Single line diagram of the Kortright Centre Microgrid

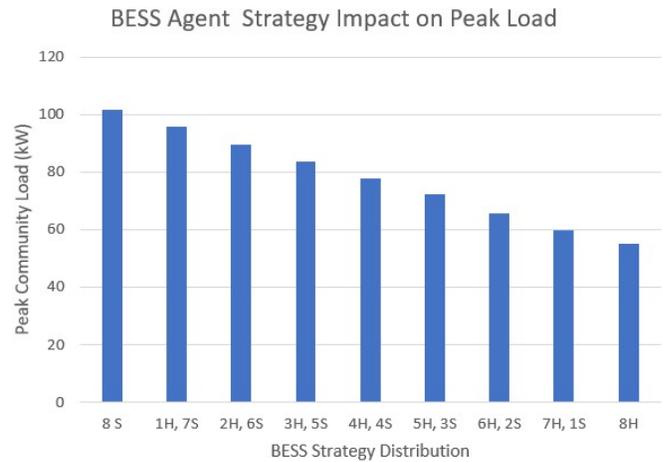

Fig. 7. Helpful BESS bidding strategies reducing community peak demand.

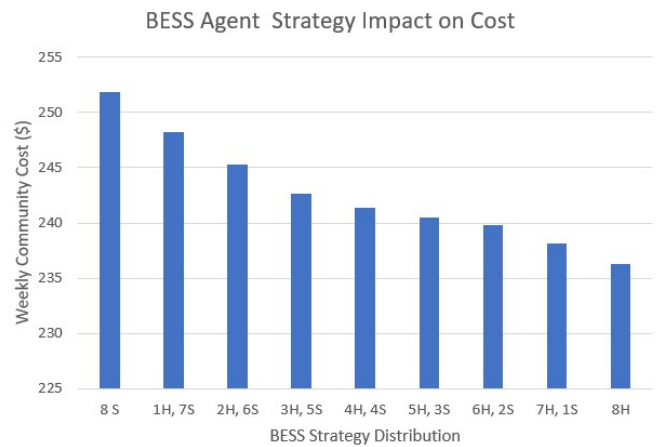

Fig. 8. Helpful BESS bidding strategies reducing community energy cost.

all smart homes.

• **Smart Contract**: Set of functions that auto-execute based on ledger data. The functions include opening a new market interval, facilitating energy bids, and executing the MCP procedure.

As such, a smart contract is triggered every time a new market interval starts, and begins to accept energy bids and measurements from the smart homes during this period. All bids are cross-verified by the peer nodes, and submitted to the ledger, where the smart contract finds the MCP of the current bid, and sends control signals back to the smart home to automatically toggle ON/OFF the participating DERs.

## V. Experimental Results

The experiments were conducted using models and real-world implementation of infrastructure at the Kortright Centre Microgrid (KCM), an initiative of the Sustainable Technologies Evaluation Program (STEP) at the Toronto and Region Conservation Authority (TRCA), located in Vaughan, Canada. As seen in Fig. 6, The KCM is equipped with 4 smart homes, 75 kW of renewable power capacity, 80 kWh of storage, and two ST systems. Two sets of experiments were carried out to validate the proposed system. The first set of results are based on a simulated, 8 home community, where the typical load profile of Smart Home 'A' was used as the baseline model for daily demand. The second set of results involved physical power transfers using the existing assets at the KCM.

### A. Simulation Results

In the simulated experiments, each smart home is assumed to have all 4 DERs installed, and a sensitivity analysis is performed to determine the impact of selfish/helpful bidding strategies for BESS and EVs on aggregate community peak demand and cost reduction based on TOU prices in Ontario. The impact of the BESS on the aforementioned objectives can be seen in Figs. 7 and 8, where the community peak demand and weekly electricity bill reduces almost linearly as the percentage of helpful BESSs increase. When comparing 8 selfish (S) BESSs against 8 helpful (H) BESSs, there is a reduction in peak load from 102 kW to 55 kW (46%), and a reduction in weekly community cost from $252 to $236 (6%). The reduction is seen primarily because a selfish BESS attempts to discharge its energy during peak times of demand, while a helpful BESS seeks to charge using available PV energy during this time.

When adding EVs to the experiments, it was found that the optimal bidding combination was 8 helpful BESSs, and a combination of 4 helpful and 4 selfish EVs. A bidding strategy of 8 helpful EVs with identical bidding curves simply creates a secondary peak during the off-peak hours of the night, which is commonly known as a rebound effect [17]. As such, a combination of 4 helpful and 4 selfish EVs helps distribute the loading impact of the EVs throughout the night. In totality, when the average community peak load is analyzed within three categories (baseline, all selfish agents, and optimal bidding), the resultant peak demands are 110 kW, 102 kW, and 47 kW, which represents a reduction of 58% from optimal bidding strategy compared to the baseline.

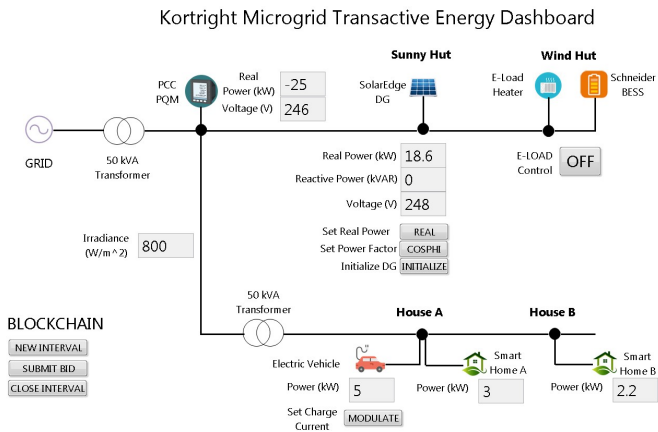

Fig. 9. DApp dashboard of the proposed platform.

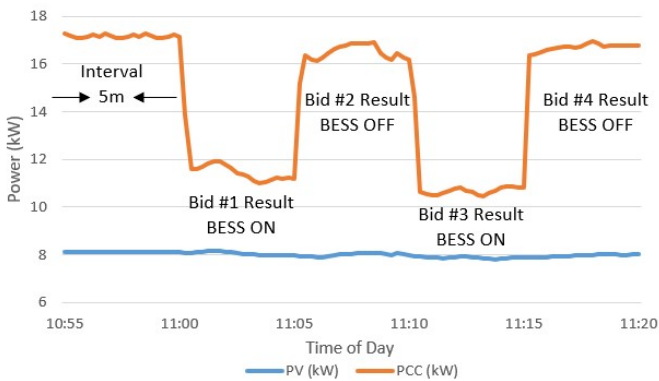

Fig. 10. Real-world demonstration of BESS operation at the microgrid.

*B. Real-world Result*

The front-end dashboard of the proposed system can be seen in Fig. 9, where the dashboard allows a user to manually open a new market interval, submit bids for any DER within the microgrid, as well as close the market interval to compute the MCP and take corresponding action for the winning DER bids. The dashboard also allows real-time visualization of power flow throughout the microgrid. A controlled experiment was conducted at the KCM using the PV of the Sunny Hut and the BESS in the Wind Hut, where a market interval was executed every 5 minutes and automated the operation of the BESS depending on its bid. The result of the experiment can be seen in Fig. 10, where the PV production and demand at the PCC of the microgrid and the main grid is plotted against time. The BESS cycles between the ON (charging) and OFF (discharging) state every consecutive market interval, as the PCC demand reduces by approximately 6 kW when the BESS is charging and increases 6 kW when the BESS is discharging.

## VI. CONCLUSION

This paper proposes a blockchain based energy trading platform that was implemented on a permissioned blockchain infrastructure and allowed energy trading amongst DERs. The proposed platform reduced the community peak load by 46% and reduced the weekly electricity bill by 6%. A real-world experiment was also conducted to validated the energy transfer between DERs within a Canadian microgrid. Future work includes expanding the algorithm to accommodate inter-community trading with the addition of ancillary services to the distribution grid.

REFERENCES

[1] M. V. Kirthiga, S. A. Daniel, and S. Gurunathan, "A methodology for transforming an existing distribution network into a sustainable autonomous micro-grid," *IEEE Transactions on Sustainable Energy*, vol. 4, no. 1, pp. 31–41, Jan 2013.

[2] U. Akram, M. Khalid, and S. Shafiq, "An improved optimal sizing methodology for future autonomous residential smart power systems," *IEEE Access*, vol. 6, pp. 5986–6000, 2018.

[3] M. Hasheminamin, V. G. Agelidis, V. Salehi, R. Teodorescu, and B. Hredzak, "Index-based assessment of voltage rise and reverse power flow phenomena in a distribution feeder under high pv penetration," *IEEE Journal of Photovoltaics*, vol. 5, no. 4, pp. 1158–1168, July 2015.

[4] W. Tushar, C. Yuen, H. Mohsenian-Rad, T. Saha, H. V. Poor, and K. L. Wood, "Transforming energy networks via peer-to-peer energy trading: The potential of game-theoretic approaches," *IEEE Signal Processing Magazine*, vol. 35, no. 4, pp. 90–111, July 2018.

[5] H. S. V. S. K. Nunna and D. Srinivasan, "Multiagent-based transactive energy framework for distribution systems with smart microgrids," *IEEE Transactions on Industrial Informatics*, vol. 13, no. 5, pp. 2241–2250, Oct 2017.

[6] Z. Li, J. Kang, R. Yu, D. Ye, Q. Deng, and Y. Zhang, "Consortium blockchain for secure energy trading in industrial internet of things," *IEEE Transactions on Industrial Informatics*, vol. 14, no. 8, pp. 3690–3700, Aug 2018.

[7] N. Z. Aitzhan and D. Svetinovic, "Security and privacy in decentralized energy trading through multi-signatures, blockchain and anonymous messaging streams," *IEEE Transactions on Dependable and Secure Computing*, vol. 15, no. 5, pp. 840–852, Sep. 2018.

[8] M. Sabounchi and J. Wei, "Towards resilient networked microgrids: Blockchain-enabled peer-to-peer electricity trading mechanism," in *2017 IEEE Conference on Energy Internet and Energy System Integration (EI2)*, Nov 2017, pp. 1–5.

[9] S. J. Pee, E. S. Kang, J. G. Song, and J. W. Jang, "Blockchain based smart energy trading platform using smart contract," in *2019 International Conference on Artificial Intelligence in Information and Communication (ICAIIC)*, Feb 2019, pp. 322–325.

[10] E. S. Kang, S. J. Pee, J. G. Song, and J. W. Jang, "A blockchain-based energy trading platform for smart homes in a microgrid," in *2018 3rd International Conference on Computer and Communication Systems (ICCCS)*, April 2018, pp. 472–476.

[11] K. J. O'Dwyer and D. Malone, "Bitcoin mining and its energy footprint," in *25th IET Irish Signals Systems Conference 2014 and 2014 China-Ireland International Conference on Information and Communications Technologies (ISSC 2014/CIICT 2014)*, June 2014, pp. 280–285.

[12] Z. Zheng, S. Xie, H. Dai, X. Chen, and H. Wang, "An overview of blockchain technology: Architecture, consensus, and future trends," in *2017 IEEE International Congress on Big Data (BigData Congress)*, June 2017, pp. 557–564.

[13] A. Brookson, "Residential energy management systems with renewables and battery energy storage," 2017.

[14] J. W. Smith, R. Dugan, and W. Sunderman, "Distribution modeling and analysis of high penetration pv," in *2011 IEEE Power and Energy Society General Meeting*, July 2011, pp. 1–7.

[15] G. Gutierrez, J. Quinonez, and G. B. Sheble, "Market clearing price discovery in a single and double-side auction market mechanisms: Linear programming solution," in *2005 IEEE Russia Power Tech*, June 2005, pp. 1–5.

[16] M. Hildmann, A. Ulbig, and G. Andersson, "Empirical analysis of the merit-order effect and the missing money problem in power markets with high res shares," *IEEE Transactions on Power Systems*, vol. 30, no. 3, pp. 1560–1570, May 2015.

[17] P. Hasanpor Divshali, B. J. Choi, H. Liang, and L. Sder, "Transactive demand side management programs in smart grids with high penetration of evs," *Energies*, vol. 10, no. 10, 2017. [Online]. Available: http://www.mdpi.com/1996-1073/10/10/1640